\documentclass[preprint,aps,showpacs]{revtex4}
\usepackage{graphicx,epsf}
\usepackage{lscape}
\usepackage[colorlinks,citecolor=blue,urlcolor=red,linkcolor=red]{hyperref}
\begin{document}

\title{{\Large{\bf Coupling constants calculation of charmed meson with kaon from QCD sum rules }}}

\author{\small M.R. seyedhabashy$^1$,
\small E. Kazemi$^2$ \footnote{e-mail: ismaeil.kazemi@gmail.com(kazemi@kazerunsfu.ac.ir)},
\small  M. Janbazi$^1$ \footnote {e-mail: mehdijanbazi@yahoo.com} and
\small N. Ghahramany$^3$ \footnote{e-mail: ghahramany@susc.ac.ir}}

\affiliation{$^1$Nuclear Science and Technology Research Institute, Tehran, Iran\\$^2$Department of Physics, Faculty of Sciences, Salman Farsi University of Kazerun, Kazerun, Iran\\$^3$Department of Physics, and Biruni Observatory, College of Science, Shiraz University, Shiraz, 71946-84795, Iran}

\begin{abstract}
In this research, the strong coupling constants of the $D^*D_s^*K$, $D_1D_{s1}K$, $D^*D_sK$ and $D_1D_{s0}^*K$ vertices are evaluated, using the three-point QCD sum rules. In order to calculate the coupling constant of each vertex, either the kaon or the charmed meson is considered as the off-shell particle. The basic $g$ parameter, in the heavy quark effective theory, is related to the coupling constants of $D^*D_s^*K$ and $D^*D_sK$. Our obtained value for $g$ parameter is $0.24\pm 0.09$, which is in good agreement with the lower limits of the other existing predictions.
\end{abstract}


\maketitle

\section{introduction}\label{s1}
Determination of the strong coupling constants helps us to better understand, the nature of the strong interactions and hadronic phenomena that are described by the nonperturbative QCD approach. QCD sum rules(QCDSR) \cite{shifman} are among the most interesting approaches at the low-energy region that have determined many successful calculations about the mesonic vertices such as $D^* D \pi$ \cite{D*Dpi1,D*Dpi2}, $D D \rho$ \cite{DDrho}, $D^* D^*\rho$ \cite{D*D*rho}, $D D J/\psi$ \cite{DDJ}, $D^* D J/\psi$ \cite{D*DJ}, $D^*D_sK$, $D^*_sD K$, $D_0 D_s K$,  $D_{s0} D K$ \cite{4Wang}, $D^*D^* P$, $D^*D V$, $D D V$ \cite{3Wang}, $D^* D^* \pi$ \cite{D*D*pi}, $D^* D^* J/\psi$ \cite{D*D*J}, $D_s D^* K$,
$D_s^* D K$ \cite{2ALozea}, $D D \omega$  \cite{DDomega}, $\phi D_{s0}^{*} D_{s0}^{*}$, $\phi D_{s} D_{s}$ , $\phi D_{s}^{*} D_{s}^{*}$, $\phi  D_{s1}
D_{s1} $ \cite{Janbazi}, $B_{s0} B K$\cite{Bs0 B K}, $B_s^* B K$\cite{Bs* B K}, $D_{s}^{*}D_{s}\phi$ \cite{Ds*Dsphi} , $D_{s}DK_0^*$, $B_{s}BK_0^*$, $D^{\ast}_{s}D K$, $B^{\ast}_{s}BK$, $D^{\ast}_{s}D K_1$, $B^{\ast}_{s}BK_1$ \cite{6Sundu}, $K^{*} K \pi$, $\phi K K$, $\phi K^{*} K^{*}$, $\rho K^{*} K^{*}$ \cite{me}, $B_1B^*\pi$, $B_1B_0\pi$, $B_1B_1\pi$, $D_1D^*\pi$, $D_1D_0\pi$ and $D_1D_1\pi$\cite{kj}.

In relativistic heavy ion collisions, the suppression of charmonium production is considered as signatures of the quark–gluon plasma \cite{matsui}. The effective SU(4) Lagrangians are applied in the low energy region to investigate the interaction of the charmonium with the hadronic medium \cite{matinyan}. The standard procedure of QCDSR is followed in our work. The theoretical [operator product expansion(OPE)] and the phenomenological(physical) contributions for the correlation function of the four proposed vertices are obtained separately, and then are equated. Higher order contributions from the OPE side and higher resonances(and continuum) from the phenomenological side are suppressed by using double Borel transformation in both sides of the equation. The numerical integration of the sum rules to estimate the coupling constant is performed. Coupling constants are functions of $q^2$(four momentum of off-shell meson) and Borel masses. The independence of the coupling constants from Borel masses are used to improve the stability of coupling constants.
 
 In the present paper, we propose the framework of the three point QCD sum rules for the calculation of the strong coupling constants of the $D^*D_s^*K$, $D_1D_{s1}K$, $D^*D_sK$ and $D_1D_{s0}^*K$ vertices. This research is reported in four sections. In section \ref{s2}, the three-point QCD sum rules framework is given in details. Our numerical findings for the coupling constants of four mentioned vertices are given in section \ref{s3}. Finally section \ref{s4} contains the conclusions. 
  
\section{three-point QCD sum rules}\label{s2}
In order to calculate the strong coupling constants of $D^*D_s^*K$, $D_1D_{s1}K$, $D^*D_sK$ and $D_1D_{s0}^*K$ in the three point QCD sum rules framework, the first step is to write the correlation functions, which for the $K$ off-shell, are given by,
\begin{eqnarray}\label{1}
\Pi_{\mu\nu}^{K}(p,p')&=&i^2 \int d^4x~d^4y~ e^{i(p^{\prime}\cdot x-p\cdot y)}{\langle}0| {\cal T}\left ( j_{\nu}^{D^*(D_{s1})}(x)~
j^{K}(0)~ j_{\mu}^{D_s^*(D_1)}(y)\right)|0{\rangle},\\
\Pi_{\mu}^{K}(p,p')&=&i^2 \int d^4x~d^4y~ e^{i(p^{\prime}\cdot x-p\cdot y)}{\langle}0| {\cal T}\left ( j_{\mu}^{D^*(D_{1})}(x)~
j^{K}(0)~ j^{D_s(D_{s0}^*)}(y)\right)|0{\rangle},
\end{eqnarray}
and for the charmed meson off-shell, are given by,
\begin{eqnarray}
\Pi_{\mu\nu}^{D_s^*(D_1)}(p,p')&=&i^2 \int d^4x~d^4y~ e^{i(p^{\prime}\cdot x-p\cdot y)}{\langle}0| {\cal T}\left ( j_{\nu}^{D^*(D_{s1})}(x)~
j_{\mu}^{D_s^*(D_1)}(0)~ j^{K}(y)\right)|0{\rangle},\\
\Pi_{\mu}^{D_s(D_{s0}^*)}(p,p')&=&i^2 \int d^4x~d^4y~ e^{i(p^{\prime}\cdot x-p\cdot y)}{\langle}0| {\cal T}\left ( j_{\mu}^{D^*(D_{1})}(x)~
j^{D_s(D_{s0}^*)}(0)~ j^{K}(y)\right)|0{\rangle}\label{1'},
\end{eqnarray}
where $j_{\mu}$ and $j$ are the pseudovector(vector) and pseudoscalar(scalar) interpolating currents which are given by quark fields as,
\begin{eqnarray}\label{2}
j^{K}(x)&=&\bar{s}(x)\gamma_5d(x),\nonumber \\
j^{D_{s0}^*}(x)&=&\bar{s}(x)Uc(x),\nonumber \\
j^{D_s}(x)&=&\bar{s}(x)\gamma_5c(x),\nonumber \\
j^{D_s^*}_{\mu}(x)&=&\bar{s}(x)\gamma_{\mu}c(x),\nonumber \\
j^{D^*}_{\mu}(x)&=&\bar{d}(x)\gamma_{\mu}c(x),\nonumber \\
j^{D_{s1}}_{\mu}(x)&=&\bar{s}(x)\gamma_{\mu}\gamma_5c(x),\nonumber \\
j^{D_{1}}_{\mu}(x)&=&\bar{d}(x)\gamma_{\mu}\gamma_5c(x).
\end{eqnarray}

where $U$ is a unitary matrix. The correlation function can be expanded in two different ways, the phenomenological and the operator product expansion(OPE). The phenomenological expression can be obtained, by inserting the three complete sets of hadronic states with the same quantum numbers between the interpolating currents, in the correlation functions, that leads to:
\begin{eqnarray}\label{3}
\Pi^{K}_{\mu\nu}&=&\frac{\langle 0|j^{D^*(D_{s1})}_{\nu}|D^*(D_{s1})(p',\epsilon')\rangle\langle
0|j^{D_s^*(D_1)}_{\mu}|D_s^*(D_1)(p,\epsilon)\rangle\langle K(q)|j^{K}|0\rangle}{(p^2-m^2_{D_s^*(D_1)})(p'^2-m^2_{D^*(D_{s1})})}
\nonumber\\&&\times\frac{\langle  D^*(D_{s1})(p',\epsilon')D_s^*(D_1)(p,\epsilon)|K(q)\rangle}{(q^2-m^2_{K})}+\mbox{....},\nonumber \\
\Pi^{D_s^*(D_1)}_{\mu\nu}&=&\frac{\langle 0|j^{D^*(D_{s1})}_{\nu}|D^*(D_{s1})(p',\epsilon')\rangle\langle
0|j^{K}|K(p)\rangle\langle D_s^*(D_1)(q)|j^{D_s^*(D_1)}_{\mu}|0\rangle}{(p^2-m^2_{K})(p'^2-m^2_{D^*(D_{s1})})}
\nonumber\\&&\times\frac{\langle  D^*(D_{s1})(p',\epsilon')K(p)|D_s^*(D_1)(q,\epsilon'')\rangle}{(q^2-m^2_{D_s^*(D_1)})}+\mbox{....},\nonumber \\
\Pi^{K}_{\mu}&=&\frac{\langle 0|j^{D^*(D_{1})}_{\mu}|D^*(D_{1})(p',\epsilon')\rangle\langle
0|j^{D_s(D_{s0}^*)}|D_s(D_{s0}^*)(p)\rangle\langle K(q)|j^{K}|0\rangle}{(p^2-m^2_{D_s(D_{s0}^*)})(p'^2-m^2_{D^*(D_{1})})}
\nonumber\\&&\times\frac{\langle  D^*(D_{1})(p',\epsilon')D_s(D_{s0}^*)(p)|K(q)\rangle}{(q^2-m^2_{K})}+\mbox{....},\nonumber \\
\Pi^{D_s(D_{s0}^*)}_{\mu}&=&\frac{\langle 0|j^{D^*(D_{1})}_{\mu}|D^*(D_{1})(p',\epsilon')\rangle\langle
0|j^{K}|K(p)\rangle\langle D_s(D_{s0}^*)(q)|j^{D_s(D_{s0}^*)}|0\rangle}{(p^2-m^2_{K})(p'^2-m^2_{D^*(D_{1})})}
\nonumber\\&&\times\frac{\langle  D^*(D_{1})(p',\epsilon')K(p)|D_s(D_{s0}^*)(q)\rangle}{(q^2-m^2_{D_s(D_{s0}^*)})}+\mbox{....},
\end{eqnarray}
where, .... denote higher resonances and continuum states. The matrix elements that appear in Eqs. (\ref{3}) can be replaced by the physical parameters with the following definitions:
\begin{eqnarray}\label{4}
\langle 0 | j^{\mathcal{P}} | \mathcal{P}(p) \rangle &=& \frac{ m^2_{\mathcal{P}} f_{\mathcal{P}}}{m_q+m_{q'}}, \nonumber \\
\langle 0 | j_{\mu}^{\mathcal{V}} | \mathcal{V}(q,\epsilon) \rangle &=& m_{\mathcal{V}}
f_{\mathcal{V}} \epsilon_{\mu}(q),\nonumber \\
\langle 0 | j^{\mathcal{S}} | \mathcal{S}(p) \rangle &=&  m^2_{\mathcal{S}} f_{\mathcal{S}},
\end{eqnarray}
where $f_\mathcal{P}$, $f_\mathcal{S}$ and $f_\mathcal{V}$ are the leptonic decay constants of pseudoscalar, scalar and vector(pseudovector) mesons respectively, $m_\mathcal{P}$, $m_\mathcal{S}$ and $m_\mathcal{V}$ are meson masses and $\epsilon _\mu$ is the polarization vector.

Generally, inserting Eq. (\ref{4}) in Eqs. (\ref{3}), leads to several lorentz structures. Theoretically, it is possible to select any structure for determination of the coupling constant. Such determined value is expected to be independent of our selected structure. Due to existing approximations in the calculations of the coupling constants, it is reasonable to select the structure, leading to more stable sum rules. Finally the phenomenological side is given by
\begin{eqnarray}\label{5}
\Pi^{K}_{\mu\nu}&=&-i g_{D_s^*(D_1) D^*(D_{s1})K}
^{K}(q^2)\frac{m_{D^*(D_{s1})} m^2_{K} m_{D_s^*(D_1)} f_{D^*(D_{s1})} f_{K} f_{D_s^*(D_1)}}{(m_u+m_s)(q^2-m^2_{K})
	(p^2-m^2_{D_s^*(D_1)})(p'^2-m_{D^*(D_{s1})}^2)}\nonumber\\&&\times(\epsilon^{ \alpha\beta \mu \nu} p_\alpha p'_\beta+...)+\mbox{higher and continuum states.}
\end{eqnarray}
\begin{eqnarray}\label{52}
\Pi^{D^*(D_{s1})}_{\mu\nu}&=&-i g_{D_s^*(D_1) D^*(D_{s1})K}
^{K}(q^2)\frac{m_{D^*(D_{s1})} m^2_{K} m_{D_s^*(D_1)} f_{D^*(D_{s1})} f_{K} f_{D_s^*(D_1)}}{(m_u+m_s)(q^2-m^2_{D^*(D_{s1})})
	(p^2-m^2_{K})(p'^2-m_{D^*(D_{s1})}^2)}\nonumber\\&&\times(\epsilon^{ \alpha\beta \mu \nu} p_\alpha p'_\beta+...)+\mbox{higher and continuum states.}
\end{eqnarray}
\begin{eqnarray}\label{53}
\Pi^{K}_{\mu}&=&- g_{D_s(D_{s0}^*) D^*(D_1)K}
^{K}(q^2)\frac{m_{D^*(D_1)} m^2_{K} [m] f_{D^*(D_1)} f_{K} f_{D_s(D_{s0}^*)}(m^2_{D_s(D_{s0}^*)}+m^2_{D^*(D_1)}-q^2)}{m^2_{D^*(D_1)}(m_u+m_s)(q^2-m^2_{K})
	(p^2-m^2_{D_s(D_{s0}^*)})(p'^2-m_{D^*(D_1)}^2)}\nonumber\\&&\times(p'_{\mu}+...)+\mbox{higher and continuum states.}
\end{eqnarray}
\begin{eqnarray}\label{54}
\Pi^{D_s(D_{s0}^*)}_{\mu}&=&-i g_{D_s(D_{s0}^*) D^*(D_1)K}
^{D_s(D_{s0}^*)}(q^2)\frac{m_{D^*(D_1)} m^2_{K} [m] f_{D^*(D_1)} f_{K} f_{D_s(D_{s0}^*)}(m^2_{K}+m^2_{D^*(D_1)}-q^2)}{m^2_{D^*(D_1)}(m_u+m_s)(q^2-m^2_{D^*(D_1)})
	(p^2-m^2_{K})(p'^2-m_{D^*(D_1)}^2)}\nonumber\\&&\times(p'_{\mu}+...)+\mbox{higher and continuum states.}
\end{eqnarray}
where $[m]=\frac{m^2_{D_s}}{m_s+m_c}$ for $D^* D_s K$ and $[m]=m_{D_{s0}^*}$ for $D_1 D_{s0}^* K$. 

The OPE side can be obtained by inserting the interpolating form of Eqs. (\ref{2}) in the correlation function of Eqs. (\ref{1}-\ref{1'}) and using the Wick's theorem in deep Euclidean region ($p^2 \rightarrow -\infty$ and $p'^2 \rightarrow -\infty$). In general the OPE process leads to:
\begin{eqnarray}\label{6}
\Pi^{OPE}=C_0+C_3 \langle \bar{q} q \rangle + C_5 \langle
\bar{q} \sigma_{\alpha\beta}T^a G^{a \alpha\beta} q \rangle +\cdots,
\end{eqnarray}
where $C_3$ and $C_5$ are the Wilson coefficients corresponding to the quark-quark and quark-gluon condensates and are the most efficient part of the nonperturbative contributions of OPE side. $C_0$ is the perturbative contribution of OPE side for each Lorentz structures,
\begin{eqnarray}\label{7}
\Pi^{per}&=&-\frac{1}{4\pi^2} \int ds \int ds'
\frac{\rho(s,s^{\prime}, q^2)}{(s-p^2)
(s^{\prime}-{p^{\prime}}^2)},
\end{eqnarray}
where $\rho(s,s',q^2)$ is the leading order of spectral density and is obtained by applying the Cutkosky's rules to the coefficients of the intended Lorentz structures,
\begin{eqnarray}\label{8}
\rho_{\mu\nu}^K=4iN_cI_0 \left[B_2(m_q+km_c)-B_1(km_{q'}-m_c)+km_c \right]
\end{eqnarray}
\begin{eqnarray}\label{82}
\rho_{\mu\nu}^{D_s^*(D_1)}=4iN_cI_0 \left[B_2(m_c+km_q)+B_1(m_q-m_{q'})+m_q \right]
\end{eqnarray}
\begin{eqnarray}\label{83}
\rho_{\mu}^K=4N_cI_0 \left[B_2(m_sm_u+km_um_c-km_cm_s-m_c^2+\Delta-\frac{u}{2})-km_cm_s-m_c^2+\frac{\Delta}{2} \right]
\end{eqnarray}
\begin{eqnarray}\label{84}
\rho_{\mu}^{D_s(D_{s0}^*)}=4N_cI_0 \left[B_2(-km_sm_u+km_um_c+m_cm_s-m_u^2+\Delta-\frac{u}{2})-km_um_s-m_u^2+\frac{\Delta}{2} \right]\nonumber\\
\end{eqnarray}
where $N_c=3$, $q=u$ and $q'=s$ for $D^* D_s^* K$, $q=s$ and $q'=u$ for $D_1 D_{s1} K$, $k=-1$ in $D^* D_s^* K(D^* D_s K)$ case and $k=1$ in $D_1 D_{s1} K(D_1 D_{s0}^* K)$ case. Also we consider the following expressions
\begin{eqnarray}\label{parameter}
I_0(s,s',q^2) &=& \frac{1}{4\lambda^\frac{1}{2}(s,s',q^2)},\nonumber \\
\lambda(a,b,c) &=& a^2+ b^2+ c^2- 2ac- 2bc- 2ac ,\nonumber \\
\Delta &=& (s+m_3^2-m_1^2),\nonumber \\
\Delta' &=& (s'+m_3^2-m_2^2),\nonumber \\
u &=& s+s'-q^2,\nonumber \\
B_1 &=& \frac{1}{\lambda(s,s',q^2)} \left [2 s' \Delta -\Delta' u\right],\nonumber \\
B_2 &=& \frac{1}{\lambda(s,s',q^2)} \left [2 s \Delta' -\Delta u\right].
\end{eqnarray}

\begin{figure}[th]
	\begin{center}
		\begin{picture}(100,00)
		\put(0,-22){ \epsfxsize=8cm \epsfbox{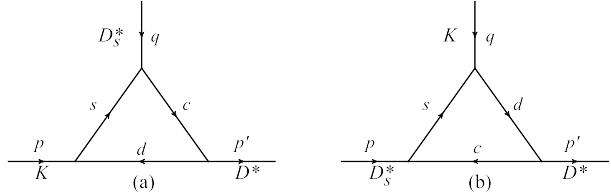} }
		\end{picture}
	\end{center}
	\vspace*{1cm} \caption{Perturbative diagrams for (a)off-shell $D_s^*$ (b)off-shell $K$.}\label{F1}
\end{figure}

\begin{figure}[th]
	\begin{center}
		\begin{picture}(100,00)
		\put(5,-20){ \epsfxsize=9cm \epsfbox{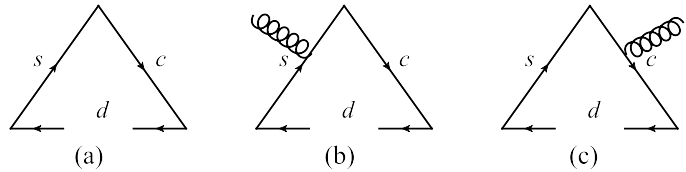} }
		\end{picture}
	\end{center}
	\vspace*{1cm} \caption{Non-Perturbative diagrams for $D_s^*$ off-shell.}\label{F22}
\end{figure}

The contribution of the perturbative term, is shown in Fig. (\ref{F1}). In Fig. (\ref{F22}), the quark-quark and quark-gluon condensates are given for the light quark spectator. 

For each vertex, the strong coupling constant can be obtained by equating the theoretical and the phenomenological representations. Let us apply the double Borel transformations to both sides of the correlation function with respect to the $p^2$ and $p'^2$, defined as follow:
\begin{eqnarray}\label{10}
{{B}}_{p^2}(M_1^2)(\frac{1}{p^2-m^2})^n=\frac{(-1)^n}{\Gamma(n)}
\frac{e^{-\frac{m^2}{M_1^2}}}{(M_1^2)^{(n-1)}}, \nonumber \\
{{B}}_{{p^{'}}^2}(M_2^2)(\frac{1}{{p^{'}}^2-m^2})^n=\frac{(-1)^n}{\Gamma(n)}
\frac{e^{-\frac{m^2}{M_2^2}}}{(M_2^2)^{(n-1)}},
\end{eqnarray}
where $M_1^2$ and $M_2^2$ are the Borel parameters. We obtained the corresponding results for the strong coupling constants as follows:
\begin{eqnarray}\label{g}
g_{D_s^*(D_1) D^*(D_{s1})K}
^{D_s^*(D_1)}(q^2)&=&i\frac{(m_u+m_s)(q^2-m^2_{D_s^*(D_{1})})}{m_{D^*(D_{s1})} m^2_{K} m_{D_s^*(D_1)} f_{D^*(D_{s1})} f_{K} f_{D_s^*(D_1)}}
~e^{\frac{m_{K}^2}{M_1^2}}
e^{\frac{m_{D^*(D_{s1})}^2}{M_2^2}} \nonumber\\& \times &\left\{-\frac{1}{4\pi^2}\int^{s^{D^*(D_{s1})}_0}_{(m_c
	+m_u(m_s))^2}ds' \int^{s^{K}_0}_{s_1} ds ~\rho_{\mu\nu}^{D_s^*(D_1)}(s,s',q^2)
e^{-\frac{s}{M_1^2}} e^{-\frac{s'}{M_2^2}}\right.\nonumber \\
&+&\left.  \frac{C_{D_s^*(D_1) D^*(D_{s1})K}
	^{D_s^*(D_1)}}{12M_1^2 M_2^2}\langle d\bar d \rangle \left(\langle s\bar s \rangle\right) \right\},\nonumber\\
g_{D_s^*(D_1) D^*(D_{s1})K}
^{K}(q^2)&=&i\frac{(m_u+m_s)(q^2-m^2_{K})}{m_{D^*(D_{s1})} m^2_{K} m_{D_s^*(D_1)} f_{D^*(D_{s1})} f_{K} f_{D_s^*(D_1)}}
 ~e^{\frac{m_{D_s^*(D_1)}^2}{M_1^2}}
e^{\frac{m_{D^*(D_{s1})}^2}{M_2^2}} \nonumber\\& \times &\left\{-\frac{1}{4\pi^2}\int^{s^{D^*(D_{s1})}_0}_{((m_c
	+m_u(m_s))^2}ds' \int^{s^{D_s^*(D_1)}_0}_{s_1} ds ~\rho_{\mu\nu}^K(s,s',q^2)
e^{-\frac{s}{M_1^2}} e^{-\frac{s'}{M_2^2}} \right\},\nonumber\\
g_{D_s(D_{s0}^*) D^*(D_{1})K}
^{D_s(D_{s0}^*)}(q^2)&=&\frac{m^2_{D^*(D_1)}(m_u+m_s)(q^2-m^2_{D_s(D_{s0}^*)})}{m_{D^*(D_{1})} m^2_{K} [m] f_{D^*(D_{s1})} f_{K} f_{D_s^*(D_1)}(m^2_{K}+m^2_{D^*(D_1)}-q^2)}
~e^{\frac{m_{K}^2}{M_1^2}}
e^{\frac{m_{D^*(D_{1})}^2}{M_2^2}} \nonumber\\& \times &\left\{-\frac{1}{4\pi^2}\int^{s^{D^*(D_{1})}_0}_{(m_c
	+m_u)^2}ds' \int^{s^{K}_0}_{s_1} ds ~\rho_{\mu}^{D_s(D_{s0}^*)}(s,s',q^2)
e^{-\frac{s}{M_1^2}} e^{-\frac{s'}{M_2^2}}\right.\nonumber \\
&+&\left.  \frac{C_{D_s(D_{s0}^*) D^*(D_{1})K}
	^{D_s(D_{s0}^*)}}{12M_1^2 M_2^2}\langle d\bar d \rangle \right\},\nonumber\\
g_{D_s(D_{s0}^*) D^*(D_{1})K}
^{K}(q^2)&=&\frac{m^2_{D^*(D_1)}(m_u+m_s)(q^2-m^2_{K})}{m_{D^*(D_{1})} m^2_{K} [m] f_{D^*(D_{s1})} f_{K} f_{D_s^*(D_1)}(m^2_{K}+m^2_{D^*(D_1)}-q^2)}
~e^{\frac{m_{D_s(D_{s0}^*)}^2}{M_1^2}}
e^{\frac{m_{D^*(D_{1})}^2}{M_2^2}} \nonumber\\& \times &\left\{-\frac{1}{4\pi^2}\int^{s^{D^*(D_{1})}_0}_{(m_c
	+m_s)^2}ds' \int^{s^{D_s(D_{s0}^*)}_0}_{s_1} ds ~\rho_{\mu}^{K}(s,s',q^2)
e^{-\frac{s}{M_1^2}} e^{-\frac{s'}{M_2^2}} \right\},
\end{eqnarray}

where the lower limits of the integrals over $s$ are:
\begin{eqnarray}\label{12}
s_1=\frac{(m_3^{2}+q^2-m_1^{2}-s')
(m_1^{2}s'-q^2m_3^{2})}{(m_1^{2}-q^2)(m_3^{2}-s')}~,
\end{eqnarray}
also, $\frac{C_{D_s^*(D_1) D^*(D_{s1})K}^{D_s^*(D_1)}}{12M_1^2 M_2^2}\langle d\bar d \rangle \left(\langle s\bar s \rangle\right)$ and $\frac{C_{D_s(D_{s0}^*) D^*(D_{1})K}^{D_s(D_{s0}^*)}}{12M_1^2 M_2^2}\langle d\bar d \rangle$ were presented as the contributions of the quark-quark and the quark-gluon condensates. The explicit expressions for them, are given in Appendix.

\section{Numerical analysis}\label{s3}
In this section, numerical analysis for the expressions of the strong coupling constant, is presented. The values of masses for quarks and mesons are given in Table \ref{mass}.                      

\begin{table}[th]
	\caption{The
		values of quark and meson masses in $\rm GeV$ \cite{PDG}.}\label{mass}
	\begin{ruledtabular}
		\begin{tabular}{ccccccccc}
			$m_s$& $m_c$&  $m_{K}$& $m_{D^{*}}$&  $m_{D_{s}^{*}}$& $m_{D_1}$& $m_{D_{s1}}$& $m_{D_s}$& $m_{D^{*}_{s0}}$\\
			\hline
			$0.14\pm0.01$& $1.26\pm0.08$& $0.49$& $2.01$& $2.11$& $2.42$& $2.46$& $1.97$& $2.32$
		\end{tabular}
	\end{ruledtabular}
\end{table}

The leptonic decay constants used in these calculations are presented in Table \ref{T1}. Also the condensate values are $\langle s\bar s\rangle=-(0.8\pm 0.2)(0.240\pm0.010 GeV)^3$ and $\langle d\bar d\rangle=-(0.240\pm0.010 GeV)^3$.

\begin{table}[th]
	\caption{The leptonic decay constants in $\rm MeV$.}\label{T1}
	\begin{ruledtabular}
		\begin{tabular}{ccccccc}
			$f_{K}$\cite{PDG,fmeson}& $f_{D^{*}}$\cite{GLWang}&  $f_{D_{s}^{*}}$\cite{Colang}& $f_{D_1}$\cite{Bazavov}& $f_{D_{s1}}$\cite{Thoma}& $f_{D_s}$\cite{Artuso}& $f_{D^{*}_{s0}}$\cite{Colang}\\
			\hline
			$156.1\pm8$& $340\pm 12$& $266\pm 32$& $294\pm27$& $ 219\pm11$& $ 225\pm20$& $ 230\pm20$
		\end{tabular}
	\end{ruledtabular}
\end{table}

The strong coupling constants presented in Eq. (\ref{g}) should not depend on the Borel mass ($M_1$ and $M_2$) variations because, the Borel parameters are not physical whereas, the strong coupling constants are physical. Therefore, we should select Borel masses region such that, the strong coupling constants remain almost stable and the results approximately close to each other. Also, for higher values of Borel parameters, the continuum and the higher states contributions may be considerable. So, in order to suppress the effect of the continuum and the higher states and simultaneously, maintain the final values of the coupling constants almost unchanged, proper limitations on the Borel parameters interval are determined.

\begin{figure}[th]
	\begin{center}
		\begin{picture}(100,00)
		\put(15,-55){ \epsfxsize=8.5cm \epsfbox{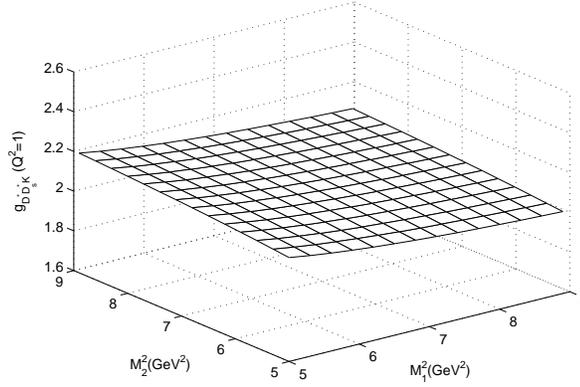} }
		\end{picture}
	\end{center}
	\vspace*{4cm} \caption{$g_{D^*D_s^*K}^K$ dependencies on Borel parameters $M_1$ and $M_2$ at $Q^2=1 GeV$.}\label{f2}
\end{figure}

The upper limits of the integrals in Eqs. (\ref{g}), are the continuum thresholds. In each vertex , if we take the initial meson mass and the final meson mass as $m$ and $m'$ respectively, then the continuum thresholds are $s_0=(m+\Delta)^2$ and $s'_0=(m'+\Delta ')^2$, and $\Delta (\Delta')$ varies between $0.4\leq \Delta (\Delta')\leq 1$\cite{kj}. By using $\Delta (\Delta')=0.5$ and fixing $Q^2=1~GeV$, we plot the dependence of the strong coupling constant $g_{D^*D_s^*K}^{K}$ on the Borel parameters, $M_1^2$ and $M_2^2$ in Fig. \ref{f2}. We found the proper interval Borel window as $5~GeV^2\leq M_1^2(M_2^2)\leq9~GeV^2$.

\begin{figure}[th]
	\begin{center}
		\begin{picture}(100,00)
		\put(5,-43){ \epsfxsize=8.5cm \epsfbox{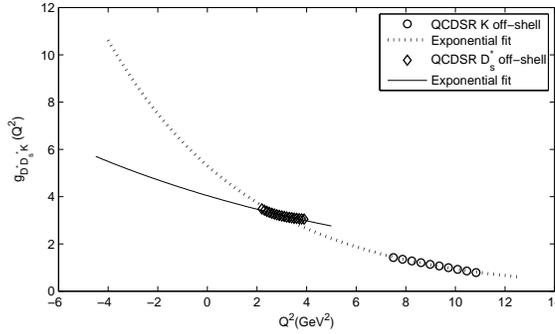} }
		\end{picture}
	\end{center}
	\vspace*{3.1cm} \caption{The strong coupling constant $g_{D^*D_s^*K}$ dependencies on $Q^2$.}\label{f3}
\end{figure}

By choosing the suitable Borel masses as $M_1=M_2=7~GeV^2$, and the inputs presented above, the $Q^2$ dependence of the numerical solution for the coupling constants can be calculated. For example, the results are presented in Fig. \ref{f3} by circles and diamonds for $g_{D^*D_s^*K}^K$ and $g_{D^*D_s^*K}^{D_s^*}$ respectively. The strong coupling constant value is defined at pole position $Q^2=-m^2_{off-shell}$.The fit function to our obtained numerical solution should be extrapolated to reach the pole position, $g(Q^2=-m^2_{off-shell})$. Our calculations show that the strong coupling constants can be fitted by the exponential fit function, as given by:
\begin{eqnarray}\label{13}
g(Q^2)=A~e^{-Q^2/B}.
\end{eqnarray}

In Fig. \ref{f3}, we show this process for $g_{D^*D_s^*K}^K$ and $g_{D^*D_s^*K}^{D_s^*}$ dependence on $Q^2$. The values of the $A$ and $B$ parameters are given in Table \ref{T2}, for two different off-shell mesons. In order to reduce the effect of approximations in choosing $\Delta (\Delta')$ and to estimate the theoretical error corresponding to these, three sets are considered. We choose $\Delta=\Delta'$ if the mass of initial and final mesons are close to each other and so $\Delta_I=0.4$, $\Delta_{II}=0.5$ and $\Delta_{III}=0.6$. For different initial and final particles, we choose $set~I=(0.6,0.4)~GeV$, $set~II=(0.7,0.5)~GeV$ and $set~III=(0.8,0.6)~GeV$, where the larger $\Delta (\Delta')$ correspond to the heavier initial(final) meson.
\begin{table}
	\caption{Parameters appearing in the fit functions for all the vertices.}\label{T2}
\begin{ruledtabular}
	\begin{tabular}{ccccccccc}
		&$\mbox{set I}$&&$\mbox{set II}$&&$\mbox{set III}$\\
		\hline
		$\mbox{Form factor}$&$A(\Delta_I)$&$B(\Delta_I)$&$A(\Delta_{II})$&$B(\Delta_{II})$&$A(\Delta_{III})$&$B(\Delta_{III})$\\
		$g^{D_s^*}_{D^*D_s^*K}$&3.60&11.09&4.04&13.07&4.46&16.34 \\
		$g^{K}_{D^*D_s^*K}$&4.99&4.27&5.31&5.77&5.45&7.61 \\
		$g^{D_1}_{D_1D_{s1}K}$&2.88&17.10&2.23&19.06&2.69&14.45   \\
		$g^{K}_{D_1D_{s1}K}$&3.69&33.16&3.90&61.56&4.28&99.47   \\
		$g^{D_s}_{D^*D_sK}$&1.46&9.06&2.03&9.38&2.63&10.24   \\
		$g^{K}_{D^*D_sK}$&3.33&4.93&3.44&6.97&3.56&9.96   \\
		$g^{D_{s0}^*}_{D_1D_{s0}^*K}$&2.71&18.80&3.72&20.54&5.14&21.01   \\
		$g^{K}_{D_1D_{s0}^*K}$&4.25&11.17&4.31&15.17&4.54&19.67   \\
		\end{tabular}
		\end{ruledtabular}
		\end{table}
	
\begin{table}[th]
	\caption{The strong coupling constants of all the vertices with the corresponding errors}\label{T3}
	\begin{ruledtabular}
		\begin{tabular}{cccccc}
			$\mbox{Coupling constant}$&&$\mbox{Coupling constant}$&\\
			\hline
			$g^{D_s^*}_{D^*D_s^*K}$&$5.64\pm1.67(\rm GeV^{-1})$&$g^{K}_{D^*D_s^*K}$&$5.48\pm1.60(\rm GeV^{-1})$\\
			$g^{D_1}_{D_1D_{s1}K}$&$3.31\pm1.23(\rm GeV^{-1})$&$g^{K}_{D_1D_{s1}K}$&$3.97\pm1.12(\rm GeV^{-1})$\\
			$g^{D_s}_{D^*D_sK}$&$3.05\pm1.31$&$g^{K}_{D^*D_sK}$&$3.57\pm0.99$\\
			$g^{D_{s0}^*}_{D_1D_{s0}^*K}$&$4.90\pm1.91$&$g^{K}_{D_1D_{s0}^*K}$&$4.44\pm1.34$\\
		\end{tabular}
	\end{ruledtabular}
\end{table}
In order to estimate the errors corresponding to the coupling constants, uncertainties in the QCD parameters such as quark and meson masses, decay constants, Borel masses, condensate values and continuum thresholds are considered in this work. To find the theoretical error corresponding to the Borel masses, the behaviour of the
coupling constants is investigated within the Borel windows, by varying the Borel mass values while fixing all other variables. The error estimation in the case of continuum thresholds is similar to the Borel masses and is calculated. The coupling constant values of each vertex and the corresponding errors are presented in Table \ref{T3}. Comparison of the results, obtained in Table \ref{T3}, we see that the method used here, to extrapolate the QCDSR results, in both cases of $D^*$ and $K$ off-shell, allow us to extract values for the coupling constants of both off-shell cases to be in good agreement with each other. The reason we chose the exponential fit instead of the linear one(which is also a good fit), is because of having such freedom to extract values of coupling constants  close to each other from both off shell cases, mentioned above. The following results for the coupling constants of the vertices studied in this work are obtained:
\begin{eqnarray}
g_{D^*D_s^*K}=5.61\pm 1.64~\rm GeV^{-1},\nonumber\\ 
g_{D_1D_{s1}K}=3.64\pm 1.19~\rm GeV^{-1},\nonumber\\
g_{D^*D_sK}=3.32\pm 1.17,\nonumber\\
g_{D_1D_{s0}^*K}=4.67\pm 1.63~.
\end{eqnarray}

Finally, we find the $g$ parameter value. The $g$ parameter is a basic value in the heavy-quark chiral effective theory and can be related to the effective coupling, by using the following formula:
\begin{eqnarray}
g_{D^*D_s^*K}=\frac{2}{f_k}g,\nonumber\\
g'_{D^*D_sK}=\frac{2\sqrt{m_{D^*}m_{D_s}}}{f_k}g.
\end{eqnarray}
where $g'_{D^*D_sK}=2g_{D^*D_sK}$. The values of $g$ parameter are presented in Table \ref{T4} and compared to the other published results. 
\begin{table}
	\begin{center}
		\begin{tabular}{|c|c|}
			\hline\hline
			$|g|$  &Reference  \\ \hline
			$0.38\pm0.08$ &\cite{HQEFT} \\      \hline
			$0.34\pm0.10$& \cite{Colangelo97} \\ \hline
			0.28& \cite{Kim01} \\ \hline
			$0.35\pm0.10$& \cite{Khodjamirian99} \\ \hline
			$0.50\pm0.02$& \cite{Melikhov99} \\ \hline
			$0.6\pm0.1$&\cite{Becirevic99} \\ \hline
			$0.59\pm0.07$& \cite{Colangelo02}  \\ \hline
			$0.27^{+0.06}_{-0.03}$& \cite{Stewart98} \\ \hline
			$0.22\pm0.10$  &  \cite{3Wang} \\ \hline
			$0.16^{+0.07}_{-0.05}$  &  \cite{4Wang} \\ \hline
			$0.24\pm 0.09$  &  This work
			\\ \hline  \hline
		\end{tabular}
	\end{center}
	\caption{ Numerical values of the  parameter $g$. }\label{T4}
\end{table}

\section{CONCLUSIONS}\label{s4}
In the present work, the three-point QCDSR is expanded to study the $D^*D_s^*K$, $D_1D_{s1}K$, $D^*D_sK$ and $D_1D_{s0}^*K$ vertices and to estimate the corresponding coupling constants. In each vertex, two off-shell meson states were considered. In order to find the corresponding coupling constant, the obtained numerical values in both off-shell meson states are averaged. These values could give useful information about the nature of the charmed meson strong interactions. The $g$ parameter is also calculated and compared to the previously found values, indicating good agreement with their lower limits.

\appendix
\begin{center}
{\Large \textbf{Appendix: The contributions of the quark-quark and
		quark-gluon condensates}}
\end{center}

In this appendix, the explicit expressions of the quark-quark and quark-gluon condensates after applying the double Borel transformations, are given by
\begin{eqnarray*}\label{condens}
C_{D_s^* D^*K}^{D_s^*}&=&\left(12M_1^2M_2^2+6m_dm_sM_2^2+3m_0^2M_2^2-6km_dm_cM_1^2-m_0^2M_1^2\right. \nonumber \\
&&\left.+3m_s^2m_d^2+3m_d^2m_c^2-3m_d^2q^2-\frac{3}{2}m_0^2m_s^2-\frac{3}{2}m_0^2m_c^2+\frac{3}{2}m_0^2q^2\right. \nonumber \\ &&\left.+6m_c^2m_d^2\frac{M_1^2}{M_2^2}+3m_c^2m_0^2\frac{M_1^2}{M_2^2}+6m_s^2m_d^2\frac{M_2^2}{M_1^2}+3m_s^2m_0^2\frac{M_2^2}{M_1^2}\right)\times e^{-\frac{m_s^2}{M_1^2}}~e^{-\frac{m_c^2}{M_2^2}}, \nonumber \\
C_{D_1 D_{s1}K}^{D_1}&=&C_{D_s^* D^*K}^{D_s^*}\mid_{s \leftrightarrow d}~,\nonumber \\
C_{D_{s0}^*D_{1}K}^{D_{s0}^*}&=&k\left(12m_sM_1^2M_2^2+6m_dm_s^2M_2^2-12m_sm_d^2M_2^2+12m_sm_0^2M_2^2+6m_dm_s^2M_1^2\right. \nonumber \\
&&\left.+6km_dm_cm_sM_1^2+6m_dm_c^2M_1^2-6m_dq^2M_1^2+6m_cm_d^2M_1^2-3m_cm_0^2M_1^2\right. \nonumber \\
&&\left.-\frac{3}{4}m_sm_d^2M_1^2+\frac{5}{8}m_sm_0^2M_1^2+12m_s^3m_d^2-6m_s^3m_0^2+12m_sm_c^2m_d^2\right. \nonumber \\
&&\left.-6m_sm_c^2m_0^2-12m_sm_d^2q^2+6m_sm_0^2q^2+3m_sm_d^2m_c^2\frac{M_1^2}{M_2^2}-\frac{3}{2}m_sm_c^2m_0^2\frac{M_1^2}{M_2^2}\right. \nonumber \\ &&\left.+3m_d^2m_s^3\frac{M_2^2}{M_1^2}-\frac{3}{2}m_s^3m_0^2\frac{M_2^2}{M_1^2}\right)\times
e^{-\frac{m_s^2}{M_1^2}}~e^{-\frac{m_c^2}{M_2^2}}=C_{D_s D^*K}^{D_s}~.
\end{eqnarray*}

\end{document}